\documentclass{aa}
\usepackage{graphicx}
\usepackage{natbib,twoopt}
\usepackage[breaklinks=true]{hyperref} %% to avoid \citeads line fills
 \bibpunct{(}{)}{;}{a}{}{,}    %% natbib format like A&A and ApJ
 \newcommandtwoopt{\citeads}[3][][]{\href{http://adsabs.harvard.edu/abs/#3}%
                                        {\citealp[#1][#2]{#3}}}
 \newcommandtwoopt{\citepads}[3][][]{\href{http://adsabs.harvard.edu/abs/#3}%
                                        {\citep[#1][#2]{#3}}}
 \newcommandtwoopt{\citetads}[3][][]{\href{http://adsabs.harvard.edu/abs/#3}%
                                        {\citet[#1][#2]{#3}}}
 \newcommandtwoopt{\citeyearads}[3][][]%
   {\href{http://adsabs.harvard.edu/abs/#3}{\citeyear[#1][#2]{#3}}}
\bibpunct{(}{)}{;}{a}{}{,}
\usepackage{latexsym}
\usepackage{subfigure}
\usepackage{amsmath}
\usepackage{amssymb}
\usepackage{amstext}
\usepackage{color}
\usepackage{psfrag}
\usepackage{txfonts}
\def\cm3{cm$^{-3}$}

\def\12{$^{12}$CO}

\usepackage{graphicx}
\usepackage{txfonts}
\begin{document} 

\title{A complete radio study of SNR~G15.4+0.1 from new GMRT observations}
\author{L. Supan\inst{1}
\and G. Castelletti\inst{1}
\and B.~C. Joshi\inst{2} 
\and M.~P. Surnis\inst{2}
\and D. Supanitsky\inst{1}
}

\offprints{G. Castelletti}
\institute {Instituto de Astronom\'{\i}a y  F\'{\i}sica del Espacio (IAFE, CONICET-UBA),
CC~67, Suc.~28, 1428 Buenos Aires, Argentina\\
             \email{lsupan@iafe.uba.ar}
\and National Centre for Radio Astrophysics (NCRA-TIFR), Post Bag No 3, Ganeshkhind, Pune - 411007, India }

 \date{Received <date>; Accepted <date>}

% \abstract{}{}{}{}{}
% 5 {} token are mandatory
 \abstract
   {}
   {The supernova remnant (SNR) G15.4+0.1 is considered to be the possible counterpart of the 
$\gamma$-ray source HESS~J1818$-$154. With the goal of getting a complete view of this remnant
and understanding the nature of the $\gamma$-ray flux, we conducted a detailed radio study that
includes the search for pulsations and a model of the broadband emission for the
SNR~G15.4+0.1/HESS~J1818$-$154 system.
}
{Low-frequency imaging at 624~MHz and pulsar observations at 624 and 1404~MHz 
towards G15.4+0.1 were carried out with the Giant 
Metrewave Radio Telescope (GMRT). 
We correlated the new radio data with observations of the source at X-ray and 
infrared wavelengths from \it XMM-Newton \rm and \it Herschel \rm observatories, respectively. 
To characterize the neutral hydrogen (HI) medium towards G15.4+0.1,  
we used data from the Southern Galactic Plane Survey. 
We modelled the spectral energy distribution (SED) using both hadronic and leptonic scenarios.
} 
   {From the combination of the new GMRT observations with existing data, we derived a 
continuum spectral index $\alpha$=$-0.62\pm0.03$ for the whole remnant. 
The local synchrotron spectra of G15.4+0.1, calculated from the combination of the GMRT data 
with  330~MHz observations  from the Very Large Array, tends to be flatter 
in the central part of the remnant, accompanying the region where the blast wave is impinging 
molecular gas. No spectral index trace was found indicating the radio counterpart to the pulsar wind nebula
proposed from X-ray observations.  In addition, the search for radio pulsations yielded negative results. 
Emission at far-infrared wavelengths is observed in the region where the SNR shock is interacting
with dense molecular clumps. 
We also identified HI features forming a shell that wraps most of the outer border of G15.4+0.1. 
Characteristic parameters were estimated for the shocked HI gas. 
We found that either a purely hadronic or leptonic model is compatible with the broadband emission 
known so far. 
}
   {}


\maketitle
\titlerunning{A complete radio study of SNR~G15.4+0.1}
\authorrunning{Supan et al.}

\section{Introduction}

The radio source \object{G15.4+0.1} was first identified as a supernova remnant (SNR) 
during the 90~cm
multiconfiguration Very Large Array (VLA) survey of the Galactic plane \citep{bro06}. This remnant 
is found to positionally coincide with 
extended emission ($\sim$8.5~arcmin in size) 
detected in the TeV $\gamma$-ray band with the High Energy Stereoscopic System (H.E.S.S.) \citep{hof11}. The TeV 
source, catalogued as \object{HESS~J1818$-$154}, is one of the faintest HESS sources detected to date. On
the basis of the morphological correspondence between the brightest hotspot of HESS~J1818$-$154 and
the inner part of G15.4+0.1, it has been proposed that a pulsar wind nebula (PWN) powered by
a still undetected rotating neutron star could be responsible for the TeV $\gamma$ rays \citep{hof11}. 

Recently, \citet{cas13} (C13)
found morphological and kinematical evidence for 
shocked CO gas in a dense molecular cloud located in the foreground of the SNR~G15.4+0.1.
Because of  the location of the cloud (partially overlapping the HESS source) and the signatures of interaction 
between the molecular gas and the remnant, the authors argued that 
the gamma radiation might be arising from  hadronic interactions 
produced by the passage of the supernova shock front through the molecular cloud.  
In that paper, based on  21~cm HI absorption 
measurements along the line of sight to G15.4+0.1,
those authors also redetermined the distance to this remnant placing it at 4.8 $\pm$ 1.0~kpc, 
instead of the 10 $\pm$ 3~kpc estimated by \citet{hof11} using 
the uncertain $\Sigma$$-$$\Delta$ relation for supernova remnants.

Later on, 
\citet{abr14} used the \it XMM-Newton \rm telescope to investigate the 
X-ray counterpart to the $\gamma$-ray source HESS~J1818$-$154.
The new X-ray image of the remnant shows extended emission in the interior of 
the radio SNR shell, in spatial coincidence with the discovered HESS source and the southern
part of the molecular cloud. 
Although the X-ray emission is fitted satisfactorily by  both the thermal and  
non-thermal models, these authors interpreted the X-rays as emission from a PWN. 
In addition, the X-ray observations lead to the detection of five point-like sources
in the region of the $\gamma$-ray emission. From their X-ray properties, only two were
thought to be galactic sources that  could be candidates for the central compact object of the SNR.
The analysis of periodicity in the signal from these sources, however, has not shown 
evidence of X-ray pulsations.

In this paper, based on new data obtained with the
Giant Metrewave Radio Telescope (GMRT),
we provide the first comprehensive analysis of the morphological and spectral radio properties of SNR~G15.4+0.1, 
which includes the results of the search for radio pulsations in the region of the source
HESS~J1818$-$154. 
In addition, we report on the study of
the spatial correspondences between the radio, infrared, and X-ray emission bands, and the analysis of
the neutral hydrogen  distribution in the environment of the remnant. On the basis of the flux measurements 
from radio to $\gamma$ ray, we discuss scenarios in which the TeV emission from this source originates from either 
hadrons interacting with dense interstellar material or leptonic emission.

\section{Observations and imaging}
\subsection{Radio continuum and neutral hydrogen observations}
The SNR~G15.4+0.1 was observed using the GMRT 
at 624~MHz on June 17 and 18, 2012 (Project code 22-015). 
The project also includes GMRT observations
of the remnant at 1420~MHz. The image at this frequency was presented for the first time 
in C13, and so it is not shown  again here. 
The data were collected in a total bandwidth of 33.33~MHz split into 512 spectral channels.
The flux density calibrator used was 3C~286, for which a flux density of 14.92~Jy was set
using the Perley-Butler 2010 VLA values\footnote{For more detailed information regarding
the flux density calibrators at VLA see the explanation of AIPS task SETJY  or
the website at http://www.nrao.edu/doc/vla/html/calib.shtml}. 
Regular observations of the source 1822$-$096 were used for phase 
and bandpass calibration. 
Reduction and imaging were carried out 
with the NRAO Astronomical Image Processing System (AIPS) 
package. The data from each day were fully reduced separately and then combined in a single
data set. The reduction of the data consisted of the removal of radio frequency interference (RFI) by hand
rather than with the automatic flagging routines, and 
bandpass and gain calibrations. After initial calibration, the data 
were averaged in terms of frequency by collapsing the bandwidth to a number of 100 spectral channels. 
For the concatenated 624~MHz data set, we employed wide-field imaging based on a pseudo-three-dimensional 
multi-facet algorithm. 
In addition, we used a multi-scale CLEAN algorithm in AIPS, with four different scale sizes. 
Several self-calibration and imaging iterations were made to obtain the final image.
The resulting GMRT image at 624~MHz has an angular resolution of 
$7^{\prime\prime}.05\,\times \,4^{\prime\prime}.55$, PA=44$^{\circ}$.42, 
and an rms noise level of 0.15~mJy~beam$^{-1}$.
We note the significantly high sensitivity (a factor of 10 better than the published image at 330~MHz) 
of this image at 624~MHz, being reported for the first time. 

On the other hand, the interstellar medium in the direction of G15.4+0.1 was investigated using data from 
the Southern Galactic Plane Survey in the 21 cm
emission line produced by HI \citep[SGPS,][]{mcc05}. 
The SGPS images, combining data from the Australia Telescope Compact Array (ATCA) and the Parkes 64 m single-dish telescope, 
are valuable in tracing large scale HI structures with an angular resolution of 
3$^{\prime}$.3 $\times$ 2$^{\prime}$.1, PA=135$^{\circ}$, and a velocity channel spacing of 0.82~km~s$^{-1}$. 

\subsection{Radio time-series observations}
\label{tsobs}
Time-series observations towards the reported centroid  
of the HESS source, $l=15^{\circ}.41$, $b=0^{\circ}.17$ \citep{hof11},  
were carried out with the 
GMRT at 1404~MHz on May 11 and 12, 2012, and at 624~MHz on June 17 and 18, 2012 
(Project code 22-015), simultaneously with the imaging observations. 
At each frequency, the GMRT was simultaneously used in two 
different modes to obtain two different time-series. In one mode, the output 
of 26 antennas were  combined in an incoherent array (IA)
with an effective beam of 43$^{\prime}$ and 24$^{\prime}$ at 624 and 1404~MHz, respectively. 
The five X-ray point sources, identified in the \it XMM-Newton \rm observations \citep{abr14}, 
all lie between 2$^{\prime}$-5$^{\prime}$ from the centre of this beam and were covered 
in the search for pulsation. 
A higher sensitivity phased array (PA) of 15 antennas, with an effective beam 
of 100$^{\prime\prime}$ and 40$^{\prime\prime}$ at 624 and 1404~MHz, respectively, and pointed at the 
centroid of the HESS position,  was used in the 
other mode. Both time-series were sampled every 61~$\mu$s 
with 512 channels across a 33.33~MHz bandpass at each frequency of 
observation. To accomplish the periodic requirement of phase calibration for both imaging and 
phased array observations, the time-series were acquired in 13  
non-contiguous integrations of about 20 minute duration each at 624 and 
1404~MHz, respectively. 
In order to assess the sensitivity of the search for radio pulsations, as well as the dispersion measures 
(DM) and the flux density estimates, we also carried out time-series observations in
direction of three known pulsars: B1642$-$03, B1937+21, and J1901$-$0906.

\subsection{X-rays observations}
With the aim of performing a multiwavelength analysis of the emission from G15.4+0.1, we reprocessed
data extracted from the on-line archive of the \it XMM-Newton \rm 
observations\footnote{\it XMM-Newton \rm Science Archive (XSA) \url{http://xmm.esac.esa.int/xsa/index.shtml}}. 
The observations in the direction of the remnant were made on October 10, 2012, during  revolution 2351 (Obs-ID: 0691390101). 

The data were reduced using the {\it XMM-Newton} Science Analysis System (SAS) 
software package Version 13.5.0. We applied standard processing to this data set to obtain 
clean event files. In order to mitigate the impact of high background flare activity, 
an appropriate screening was applied, extracting a light curve for photons above 10~keV, 
for the entire field of view. We selected events with {\tt FLAG} $=0$ and {\tt PATTERN} $\leqslant 12$ 
for both EPIC MOS1 and MOS2 cameras \citep{tur01}. 
The resulting total effective exposure time of observation with the MOS1 and MOS2 cameras was 30~ks.

In order to reveal the diffuse emission, point-like sources within the field of view were 
extracted by using the {\tt dmfilth} tool of the CIAO (v. 4.6) reduction 
package\footnote{More information available at \url{http://cxc.harvard.edu/ciao/}} with {\tt POISSON} statistics to fill the excluded regions. Finally, a single combined image in the energy range 1-8~keV from both cameras was obtained, which was then smoothed with a Gaussian kernel to a resolution of 
$\sim$1$^{\prime}$.4.

\section{Results}
\subsection{Total intensity image of G15.4+0.1 at 624~MHz}
\begin{figure}[!ht]
\centering
\includegraphics[width=8cm,clip]{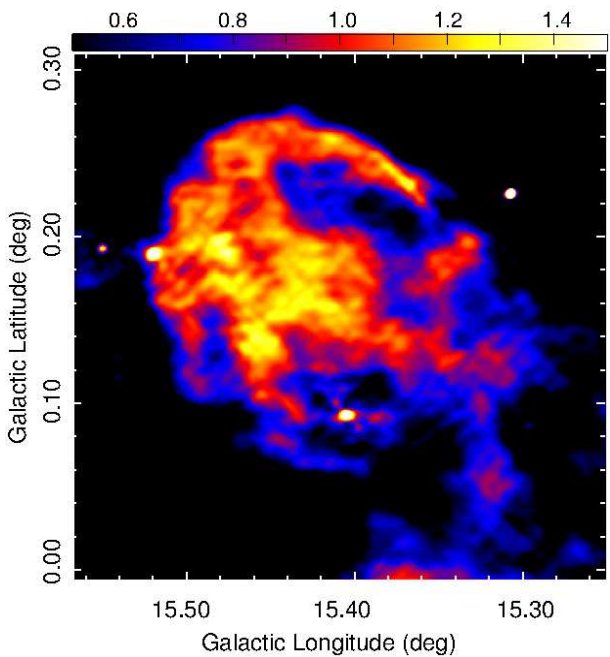}
\caption{Total intensity image of SNR~G15.5+0.1 at 624~MHz obtained with the GMRT. 
It does not include primary beam correction.
The image was smoothed to an angular resolution of $10^{\prime\prime}$, resulting in an rms noise level
of  0.48~mJy~beam$^{-1}$. The unit of the values in the colour bar is mJy~beam$^{-1}$.
}
 \label{fig1}
\end{figure}

Figure~\ref{fig1} shows the new image of SNR~G15.4+0.1 obtained at 624~MHz using the GMRT. It constitutes
the first ever high-resolution and sensitivity image of this source at low radio frequencies. 
The shortest baseline available in the GMRT observations at 624~MHz is 
$\sim$197$\lambda$, which  
implies that spatial scales of the G15.4+0.1 emission up to $\sim$17.4~arcmin
are well sampled in the resulting image. 
The radio emission from this remnant has the appearance of an irregular shell 
somewhat elongated in the 
Galactic north-east to south-west\footnote{Although we do not utilize the equatorial coordinate system, we refer 
hereafter to north and south to set out locations around the Galactic plane.} direction 
(i.e. from  upper left to lower right of Fig.~\ref{fig1}) 
with an average 
diameter of  14.5~arcmin (i.e. $\sim$20~pc at the distance of 4.8~kpc derived in C13).
The same overall total intensity morphology is also observed in the images of this remnant at 330 and 
1420~MHz reported by \citet{bro06} and in C13 based on VLA and GMRT observations, respectively.
At 624~MHz, the emission is brighter and patchy along the Galactic northwestern  
rim and from the northeastern to the central part of the radio shell, overlapping the region where 
shocked CO gas has been identified (C13); see also Fig.~\ref{multiwave} in current work.
The bright spot at $l=15^{\circ}.52$, $b=0^{\circ}.19$ corresponds to the 624~MHz radio counterpart
of the bipolar bubble HII region G015.520+0.188 \citep{and11}, for which \citet{sup14} 
have determined a distance of $3.0 \pm 0.3$~kpc based on the new GMRT observations and HI emission line data.

The radio emission becomes fainter and uniform towards the southern part of G15.4+0.1. Unlike the north edge, 
no sharp boundary is apparent in the south limb of the remnant. A gap in the radio emission is also evident along 
the western edge of the remnant.

To accurately determine the flux density integrated over the whole 624~MHz emission from G15.4+0.1, 
we subtracted the contribution of the aforementioned HII region and the bright knot located at 
$l$$\sim$$15^{\circ}.48$, $b$$\sim$$0^{\circ}.20$, which is likely to be 
an extragalactic source, previously identified as NVSS~J181806$-$152321 
(NED database)\footnote{NED (the NASA/IPAC Extragalactic Database) 
is operated by the Jet Propulsion Laboratory, California Institute of Technology, 
under contract with the National Aeronautics and Space Administration.}. The integrated flux    
density for G15.4+0.1 based on the new GMRT observations at 624~MHz is  $8.0 \pm 1.1$~Jy using the
\citet{per13} flux density scale. The error in our estimate
reflects the uncertainty both in the determination of background emission and in the selection of the area
over which we integrate the remnant flux.

\subsection{Pulsar search}
\label{psrsrc}
\begin{figure*}[!ht]
\centering
\includegraphics[width=13cm]{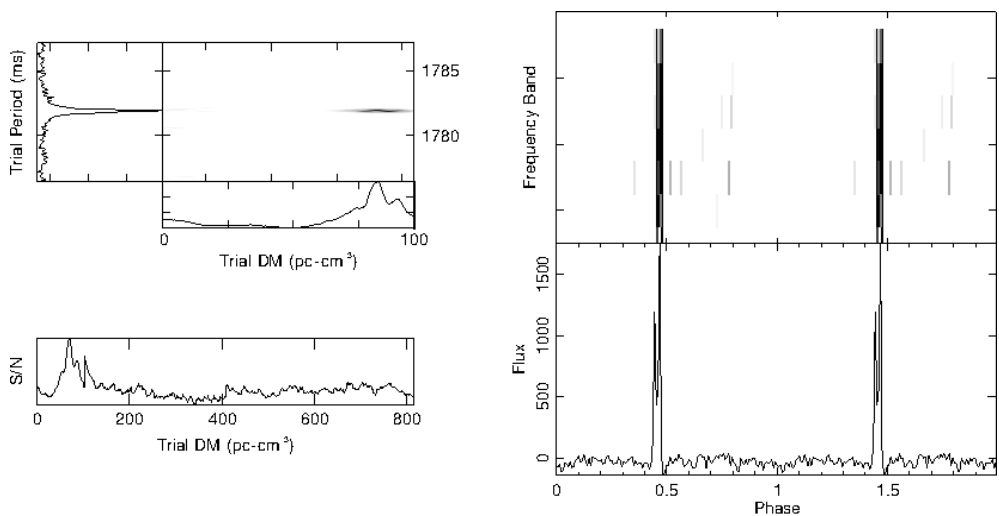}
\caption{Visual representation of a pulsar search candidate. The plot shows 
the detection of the known pulsar PSR~J1901$-$0906,
following the procedure described in the text. 
The bottom left panel  shows the peak in the S/N 
curve over a large range of dispersion measures (DM) values, whereas the 
top left panel indicates the peak in the S/N with a refined 
period and DM. The top right plot  shows a greyscale representation 
of intensity as a function of pulse phase in abscissa and the bandpass 
in ordinate at the refined period and DM, while the bottom right 
plot shows the averaged profile of the pulsar. Two rotations are 
shown for clarity in this plot. 
This representation helps distinguish mistaken RFI candidates from genuine candidate pulsars.}
\label{figviewer}
\end{figure*}

The time-series data for both the incoherent and coherent array 
for each frequency were analyzed using 
SIGPROC\footnote{http://sigproc.sourceforge.net} pulsar data analysis 
software using a 16 core 32-node High Performance Computing cluster (HPC) 
at the National Centre for Radio Astrophysics (NCRA, India). First, the channels 
affected by RFI were flagged from the 
data. The usable bandwidth of the observations was somewhat less 
than 33.33 MHz owing to the presence of narrowband RFI channels, 
which were removed, and was different for incoherent and phased 
array observations. 
These were then dedispersed to  1792 trial dispersion 
measures\footnote{The integrated electron column density along the line of sight in units 
of ~pc~cm$^{-3}$.} (DMs) covering a range of 0-815~pc~cm$^{-3}$ at 624~MHz. The DMs 
were spaced more coarsely at 1404~MHz, owing to a small amount of 
dispersion smear across each channel. Data were dedispersed to 
 360 trial DMs at this frequency covering 0-1200~pc~cm$^{-3}$. 
A harmonic search was performed on the dedispersed time-series for each DM. 
Because of the non-contiguous nature of the time-series, an 
incoherent addition of spectra from individual time-series segments  
was carried out  before 
the harmonic search. Known RFI periodicities, such as power-line 
frequencies at 50 and 100~Hz, were excised from the resultant 
spectra before the harmonic search. The resultant candidate 
periodicities were carefully examined using a visual representation 
of the search results as shown in Fig.~\ref{figviewer}. The 
search typically produced  60 to 70 candidates 
at each frequency 
for the phased array observations, while the number of candidates 
for the incoherent array were larger. No pulsations above a threshold 
signal-to-noise ratio (S/N) of 8 were detected in either phased array 
or incoherent array observations at both frequencies. 

\begin{figure}[!ht]
\centering
\includegraphics[width=8cm]{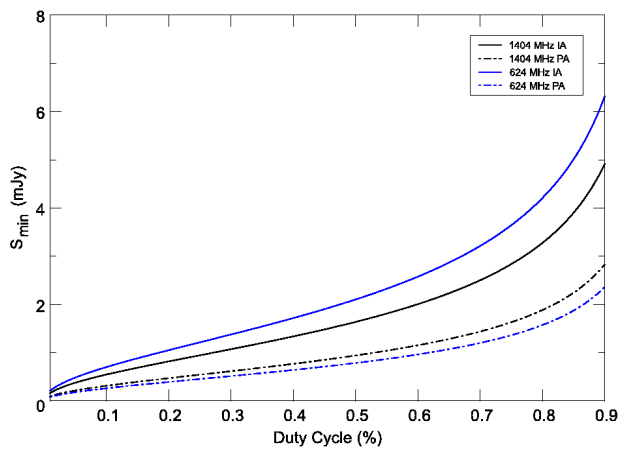}
\caption{Sensitivity curves for the pulsational search. The plot 
depicts the 8 standard deviation upper limits obtained from 
our observations for incoherent array at 1404~MHz (solid black), 
phased array at 1404~MHz (dashed-dot black), incoherent array at 
624~MHz (solid blue), and phased array at 624~MHz (dashed-dot blue) 
as a function of duty cycle for a possible pulsar.}
\label{figsens}
\end{figure}

The sensitivity of the pulsation search was assessed using 
observations of the three known pulsars, PSRs B1642$-$03, 
B1937+21, and J1901$-$0906, as mentioned in Section \ref{tsobs}. 
A procedure similar to that outlined in \citet{cjp12} was used 
for this purpose. The sensitivity of pulsation search depends 
on the duty cycle (ratio of pulse width to pulsar period) of 
the pulsar. As SNR~G15.4+0.1 
lies in the direction of the Galactic centre, any associated pulsar 
is likely to be scatter-broadened, particularly at 624~MHz, which 
affects the duty cycle of the pulsar. For example, 
using NE2001 models \citep{cor02} with a 
distance of 4.8~kpc to the SNR (C13), we get an estimate of 
temporal broadening of about 13~ms at 624~MHz, whereas this 
value is about 300~$\mu$s at 1404~MHz. Hence, we have plotted our 
sensitivity curves as a function of duty cycle in Fig.~\ref{figsens} 
to provide a more meaningful comparison in the presence of 
temporal broadening. The individual curves correspond to the incoherent 
array and the phased array observations at 624 and 1404~MHz, respectively. 
These curves provide 8 standard deviation upper limits, for pulsed 
emission with 10\% duty cycle, of 250 and 300~$\mu$Jy in the 
phased array observations 
at 624 and 1404~MHz, respectively, from any 
putative pulsar associated with the SNR~G15.4+0.1/HESS~J1818$-$154 system. 
Upper limits of 700 and 500~$\mu$Jy in the incoherent array observations 
at 624 and 1404~MHz, respectively, are also implied for the 
five identified X-ray point sources within the extent of the 
SNR.  We note that the incoherent array beam covers the entire remnant.  

\subsection{A multiwavelength view of SNR~G15.4+0.1}
\begin{figure*}[!ht]
\centering
\includegraphics[width=13cm, clip]{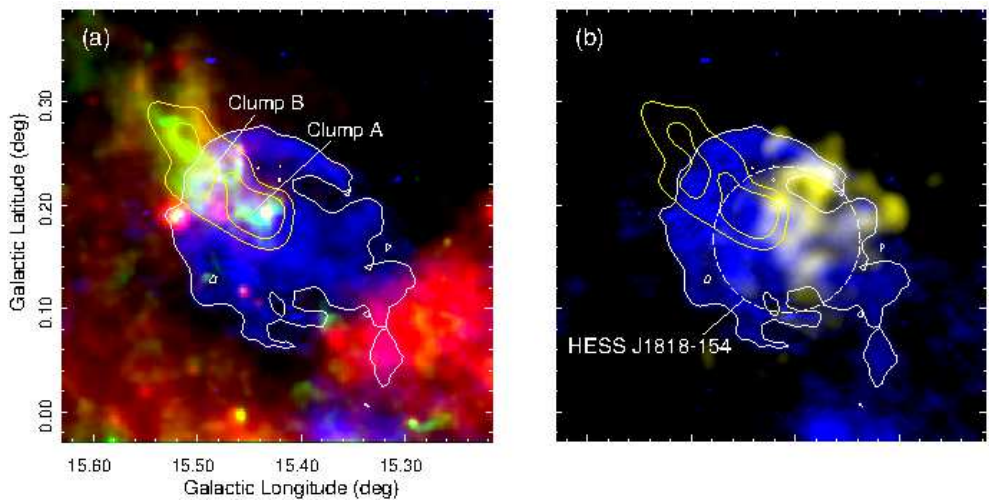}
\caption{Multiwavelength view of SNR~G15.4+0.1. \bf a) \rm A color composite image showing the 
new radio image at 624~MHz of G15.4+0.1 (in blue) and FIR emission as observed 
by \it Herschel \rm at 70~$\mu$m (in red), and 250~$\mu$m (in green);
\bf b) \rm Comparison of the radio continuum emission at 624~MHz of the SNR
(in blue) and the X-ray emission (in yellow) as observed by \it XMM-Newton \rm in the 
1-8~keV energy range. The dashed white circle indicates the size and location of the TeV 
source HESS~J1818$-$154. 
In both panels, the yellow contours delineate the $^{13}$CO line emission in the 46-50.3~km~s$^{-1}$ range 
(C13).}
 \label{multiwave}
\end{figure*}

We have explored the spatial co-existence of synchrotron radio features and
sites of emission at infrared and X-ray wavelengths. In order to compare radio 
and far-infrared (FIR) morphologies, we made use of archival data at 70 and 250~$\mu$m
acquired using the SPIRE instrument on \it Herschel Space 
Observatory\footnote{Herschel is an ESA space observatory with science instruments 
provided by European-led Principal Investigator consortia and with important participation from NASA.}\rm
\citep{gri10}. 
Far-infrared images are a very useful tool with which to delineate shock heated dust, particularly
in regions where the SNR shock front is hitting dense interstellar medium.
The spatial correspondence between the new radio image at 624~MHz and the
FIR images is shown in Fig.~\ref{multiwave}a; 
yellow contours are included to help locate the molecular cloud discovered by C13.
To perform the radio/IR comparison, the 70~$\mu$m 
data were convolved to the same angular resolution as the 250~$\mu$m map 
(18$^{\prime\prime}$). From Fig.~\ref{multiwave}a, the correlation between the CO molecular cloud
and the spatial distribution of the emission at IR wavelengths is clearly evident.
Indeed, the elongated structures that appear in the FIR images is
completely bounded within the dense CO structure detected at velocities between 
46 and 50.3~km~s$^{-1}$ (C13). Moreover, significant enhancements in the dust emission at the two 
wavebands are located inside the two molecular clumps (referred to in C13 as clumps A and B). 
Such coupling between the dust emission in the FIR and the CO line emission provides
additional support to the presence of dense molecular gas associated with the SNR. 
We thus interpret the infrared emission as marking the SNR-molecular cloud interface in G15.4+0.1. 
Evidence of a similar situation has also been observed in the northern shock front of Cas~A,
in the region where the remnant encounters  a density enhancement in the interstellar medium (ISM) \citep{hin04}.

To obtain a more complete description of the infrared distribution in G15.4+0.1, we 
also used data at 3.6, 4.5, 5.8, and 8~$\mu$m taken with the IRAC camera 
aboard the \it Spitzer Space Telescope \rm along with MIPSGAL images at 24~$\mu$m.
Aside from the striking correspondence observed at 
FIR wavebands with the CO cloud, no discernible emission was detected in the 
near- or mid-infrared wavebands, which correlates with either the radio surface brightness 
or the proposed PWN in X-rays (the comparisons are not shown here).
This result contrasts with that observed in several PWNe in the Galaxy, for which there is a 
spatial agreement between synchrotron radio, X-rays, and 
near-infrared emission distribution
(see e.g. the multiwavelength view of the PWNe associated with SNRs~3C~58, \citet{sla08}; 
Crab \citet{san09}; G21.5$-$0.9, \citet{zaj12}; and G292.0+1.8, \citet{zha13}).

The comparison of the emission in radio and X-ray bands was first presented by 
\citet{abr14} using the image of the remnant taken from 
 the 330~MHz VLA survey of the Galactic plane \citep[GPS,][]{bro05gps} and \it XMM-Newton \rm observations. 
Here, we redo this analysis in the light of the new GMRT data by combining the synchrotron emission at 
624~MHz from the source, which shows
a highly structured interior compared to 
that observed at the lower radio frequency, with \it XMM-Newton \rm reprocessed data obtained from
X-ray energies between 1.0 and 8.0~keV. The radio/X-ray morphological correlation 
is shown in Fig.~\ref{multiwave}b. Contours for the molecular gas interacting
with the SNR (C13) together with the region where $\gamma$-ray emission was observed
are also presented in the radio/X-ray comparison. 
Fig.~\ref{multiwave}b reveals little correspondence between both emitting plasmas.
The X-rays associated with G15.4+0.1 do not show evidence of a limb-brightened X-ray shell, but most
of the X-ray emission appears broadly contained within the radio rims of G15.4+0.1. 
Moreover, the comparison shows X-ray emission
extending slightly beyond the projected radio boundaries of the remnant.
In addition, weak and diffuse X-ray emission is present filling the spatial gap 
observed in radio all along the western limb of the remnant. 

Figure~\ref{multiwave}b also demonstrates 
a significant lack of  correspondence between the X-rays and
the molecular material, 
delineated by yellow contours. This result may not be surprising if
obscuration due to the molecular cloud detected 
between the observer and the remnant occurs. 
On the contrary, the morphology in X-rays overlaps part of the $\gamma$-ray emission area delineated by
the white circle in Fig.~\ref{multiwave}b. 
Such a X-ray/$\gamma$ ray correlation led \citet{abr14} to  the conclusion that the X-ray emission 
is the counterpart to the TeV source originating  in a 
pulsar wind nebula of a yet undiscovered pulsar.
This interpretation, however, may not have a robust observational support since
the authors remark that both thermal and non-thermal models are equally satisfactory to fit the spectrum
of the diffuse X-rays in G15.4+0.1,
which made somewhat uncertain the true 
origin of the detected X-ray emission.
Finally, we found no radio counterparts corresponding to the five X-rays point sources  
detected using \it XMM-Newton \rm observations at the location of the emission in $\gamma$ rays 
\citep{abr14}.  

\section{Radio spectral analysis in SNR~G15.4+0.1}
\subsection{The integrated spectrum}
\begin{figure}[!ht]
\centering
\includegraphics[width=8cm]{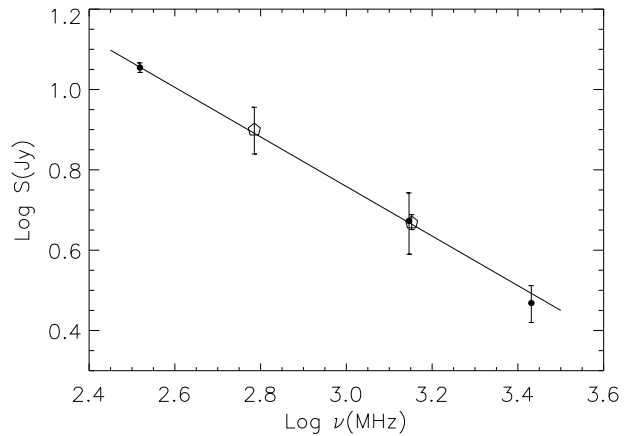}
\caption{Integrated radio continuum spectrum of SNR~G15.4+0.1. The open symbols correspond to 
GMRT flux densities measurements at 624 and 1404~MHz. The rest of the data were taken
from \citet{bro06}. All the data were tied to a common flux density scale \citep{per13}. 
The solid line shows the best fit obtained with a single power law to all the values, 
which yields a global spectral index $\alpha$=$-0.62\pm0.03$.}
 \label{spectrum}
\end{figure}

To determine the global spectral index ($S \propto \nu^{\alpha}$) of G15.4+0.1,
we have used the new flux density estimate at 624~MHz together with that obtained from GMRT data at 
1420~MHz (C13) and measurements at 330, 1400, and 2700~MHz as determined from observations carried 
out with the VLA and Effelsberg telescopes \citep{bro06}.
In Table~\ref{integrated} we summarized the flux densities integrated  over the whole extension of 
this remnant at the mentioned frequencies.
The listed values constitute all the measurements published up to present for the source.
A correction factor was applied to set all the values to the \citet{per13} absolute flux density scale. 
In particular, the correction to the flux density estimates at 330 and 624~MHz 
was made using the low frequency extension proposed by those authors. 
Figure~\ref{spectrum} shows the resulting radio continuum spectrum of the remnant; 
the determinations from GMRT observations are plotted as open pentagons symbols. 
A weighted fit to all the available flux densities
yields  a spectral index $\alpha$=$-0.62\pm0.03$, 
which is compatible with the value derived previously by \citet{bro06}.
This spectral behaviour is consistent with that measured in other shell type SNRs \citep{kot06}.

\begin{table}[h!]
\caption{Integrated flux densities for G15.4+0.1.}\label{fluxes}
\centering
\begin{tabular}{l c c}\hline\hline
 Frequency  & Scaled           & References    \\ 
      (MHz) & flux density (Jy)  \\ \hline
      330       & $11.3 \pm 0.3$       & \citet{bro06}     \\
       624      & $8.0 \pm 1.1$        & This work         \\
       1400     & $4.7 \pm 0.8$        & \citet{bro06}     \\
      1420      & $4.7 \pm 0.2$        & C13               \\
      2700      & $2.9 \pm 0.3$        & \citet{bro06}     \\
\hline
\label{integrated}
\end{tabular}
\end{table}

\subsection{Spatial variations in the radio spectral index}
\begin{figure*}[!ht]
\centering
\includegraphics[width=13cm]{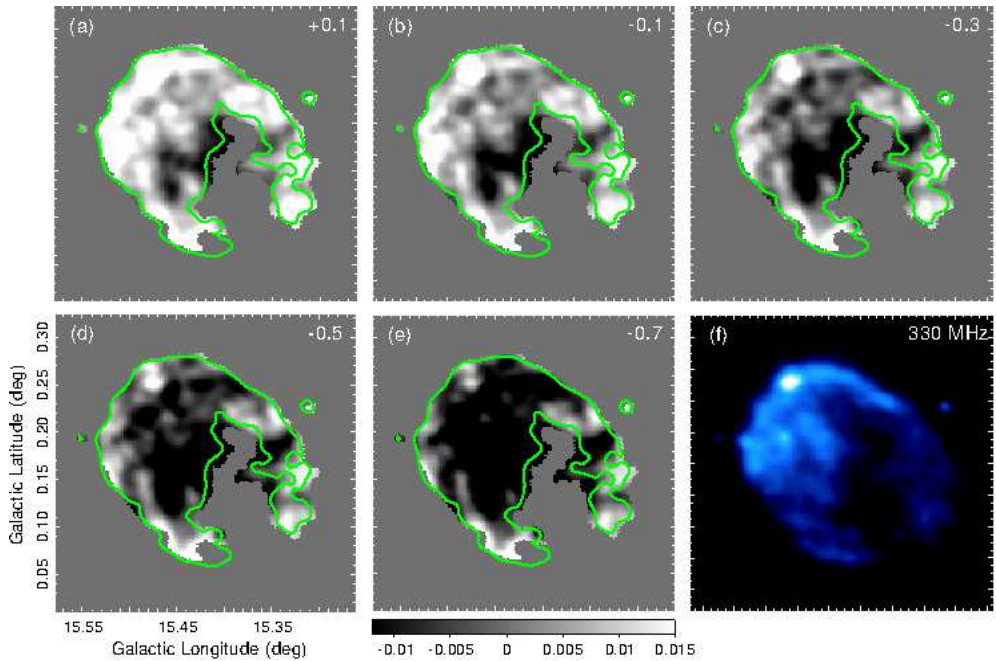}
\caption{Tomographic images for SNR~G15.4+0.1 constructed from data at 330 and 624~MHz. The data
at both frequencies were matched to the same uv-coverage. The radio spectral index $\alpha_{\mathrm{t}}$ 
is indicated at the top right corner  of the panels. 
The grey colour bar in the middle column represents the scale of the residuals 
components, which is the same for all tomographies.  
In the greyscale representation, 
spectral components for which $\alpha$ is steeper than the assumed $\alpha_{\mathrm{t}}$ are 
shown as light regions (positive residuals), while flatter components are seen
as dark features (negative residuals). A contour delineating the radio continuum emission 
at 330~MHz is included for reference. The total intensity image of G15.4+0.1 at 330~MHz is 
plotted in the bottom right frame \citep{bro06}. }
\label{tomo}
\end{figure*}

We have investigated the spectral changes as a function of the position within the remnant by using 
the image at 624~MHz and the one at 330~MHz (see panel (f) in Fig.~\ref{tomo}) kindly provided by C.~Brogan 
\citep{bro06}. The analysis was done by constructing tomographic maps between them \citep{kat97}. 

Basically, the procedure involves the generation of a series of 
residual images $S_{\mathrm{t}}$ according to the expression
$S_{\mathrm{t}} = S_{330} - \left(\frac{330}{624}\right)^{\alpha_{\mathrm{t}}} S_{624}$,
where $S_{330}$ and $S_{624}$ are the maps at 330 and 624~MHz, respectively, and $\alpha_{\mathrm{t}}$ 
represents a trial spectral index value. If the assumed spectral index coincides with 
the ``true'' value, one should expect the 
$S_{\mathrm{t}}$ image to be equal to zero.
On the contrary, features in the residual map that  
appear positive (negative) will be
associated with local spectral indices that are steeper (flatter) than the assumed $\alpha_{\mathrm{t}}$ value.
The technique has the advantage of delineating spectral features that can 
overlap along the line of sight while avoiding any zero level dependence.

In Fig.~\ref{tomo}, we reproduce the gallery of tomographic maps between 330 and 624~MHz. To
construct them,  the radio images were 
clipped at the 3$\sigma$ level of their respective noise levels.
In order to obtain a reliable determination of the spectral properties, 
the range of the spatial scales measured at 624~MHz was matched in the \it uv\rm-coverage to 
that of the 330~MHz data. Additionally, the images were aligned and interpolated to identical 
projections to avoid positional offsets. The $\alpha_{\mathrm{t}}$ value shown in the maps
varies from +0.1 to $-$0.7. 
For the sake of illustration the 330~MHz total intensity image of G15.4+0.1 is
shown in the bottom right frame in Fig.~\ref{tomo}.

Spatial variations in the spectral index are recognizable across the synchrotron emission from G15.4+0.1.
The central part of the remnant shows components of spectral index flatter ($\alpha$ varying 
between $\sim$+0.1 and $-$0.5) than the values in the ridge. 
We note that the very flat spectral component observed as a dark area in the 
southeastern border of the remnant ($l$$\sim$15$^{\circ}$.40, $b$$\sim$0$^{\circ}$.15),  
within which the spectral index appears to have positive values 
(see panel corresponding to $\alpha_{\mathrm{t}}$=$+$0.1 in Fig.~\ref{tomo}), corresponds to
a bordering clipped area in the map and should not be considered  a real spectral component
owing to the decrease in the signal-to-noise ratio observed in this region in the VLA 330~MHz image
in comparison with the image at 624~MHz.
In general, the central flattening traced in the 330/624~MHz comparison  
matches the zone where the interaction with the discovered CO molecular cloud
is taking place (see Fig.~\ref{multiwave}). 
Variations in the spectrum are generally expected in regions where 
the SN shock is impacting high density interstellar material as a consequence of the density 
increase and the deceleration of the shock in the interacting regions.
For instance, spectral flattening were found between 74 and 330~MHz 
for the two well-known SNRs/molecular clouds systems  
marking the ionized interface of molecular clouds interacting with the shock front in 
3C~391 \citep{bro05} and IC~443 \citep{cas11}. 

The tomographies also show a conspicuous bullet-like feature standing out as a bright 
spectral component (i.e. positive in the $S_ {\mathrm{t}}$ maps) located at the 
northeast part of the shell ($l$$\sim$15$^{\circ}$.48, $b$$\sim$0$^{\circ}$.25). It is characterized by a 
spectral index even steeper than $-$0.7 because it appears as a positive residual  in all tomographic images. 
We found no catalogued extragalactic objects at the position of this feature. 

Along the westernmost edge (left side) of the SNR, there is an indication 
of a positive correlation between the local spectral index and 
the total radio intensity features in the sense that  the bright ridge 
tends to be associated with steeper indices. A similar situation was found in young SNRs  
\citep[see e.g. the analysis of SNRs G39.2$-$0.3 and G41.1$-$0.3 in][]{and93}. In those cases, 
it was suggested that the radio brightness and spectra are being regulated by different mechanisms.  
In addition to this behaviour, 
most of the structures in the tomographies traced at $\alpha_{\mathrm{t}}$=$-$0.5 and
$\alpha_{\mathrm{t}}$=$-$0.7 disappear against the background indicating that the spectral
behaviour can be characterized by a mean spectral index $\sim$$-$0.6, which is consistent with 
the integrated spectral index of this SNR derived by fitting the total flux densities from 330 to 
2700~MHz (see Sect.~4.1). 
Overall, the observed radio continuum spectral distribution across G15.4+0.1 is compatible
with what is expected in the framework of diffusive shock acceleration models.

\section{Neutral gas distribution around G15.4+0.1}
\subsection{HI brightness morphology and physical parameters}
We explored the interstellar medium around G15.4+0.1 in the 21~cm line to distinguish some morphological
correlations between the distribution of the neutral gas and the remnant that might account for the SNR 
shock effects on the surrounding medium. 

\begin{figure*}[!ht]
\centering
\includegraphics[width=13cm, clip]{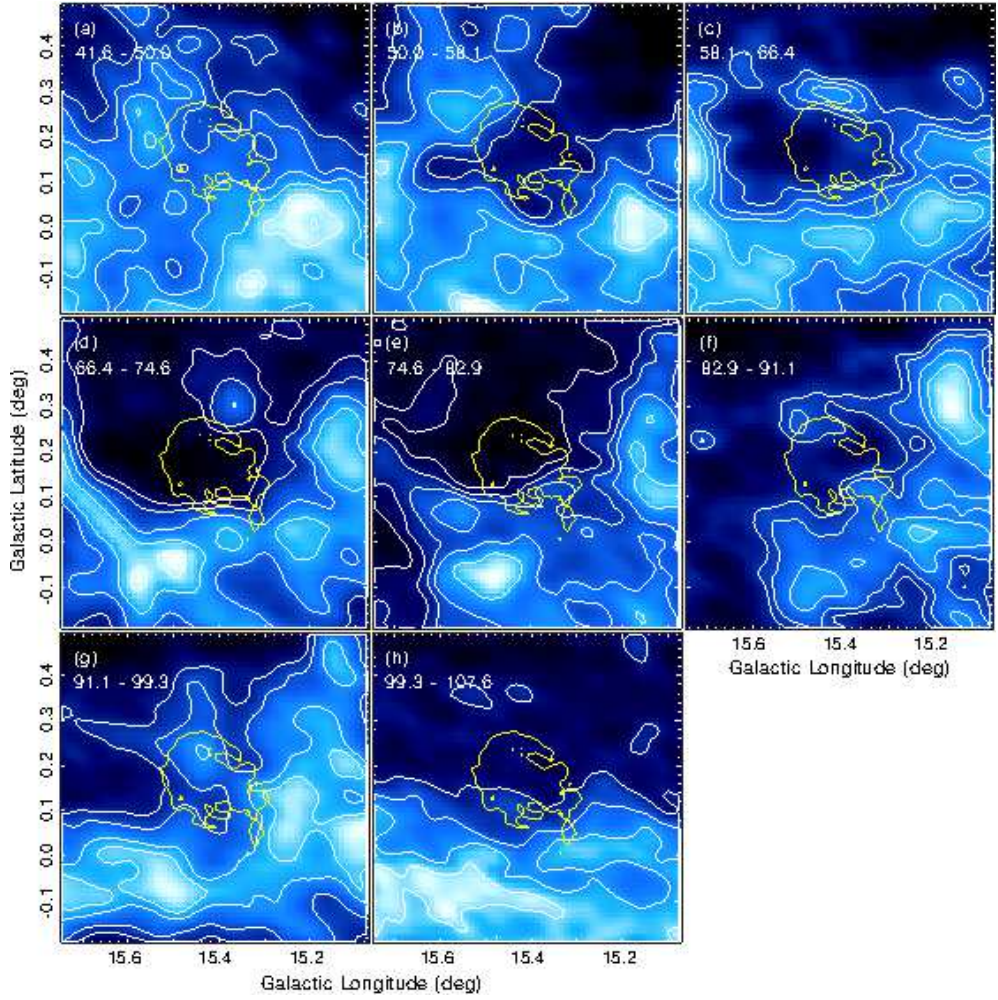}
\caption{Images of 21~cm line emission in the direction of SNR~G15.4+0.1. Each map corresponds to
the integration of the neutral gas every 8.2~km~s$^{-1}$. The range of velocities is indicated 
at the top right corner of the panels. The colour display was kept for all images between 
633 and 1210~K~km~s$^{-1}$. The 0.7~mJy~beam$^{-1}$ contour
from the 35$^{\prime\prime}$ resolution 624~MHz image is included in each panel to facilitate the comparison
between the radio continuum emission from the remnant and its surroundings.}
\label{HImaps}
\end{figure*}

Figure~\ref{HImaps} includes velocity channels maps 
from the SGPS data cube covering an interval from 41.6 to $\sim$108~km~s$^{-1}$.  
All the velocities  are measured with respect to the local standard of rest (LSR). 
Each frame was obtained by integrating the data cube every 8.2~km~s$^ {-1}$. 
An appropriate background level equal to the mean value of each integrated image was subtracted from every 
map. A contour line delineating the outer boundary of the 
radio continuum emission at 624~MHz 
from G15.4+0.1 was overlaid to the HI distribution for ease of comparison.

The distribution of the neutral gas in the velocity range from 41.6 to 50~km~s$^{-1}$ only shows 
bright clouds spread over the field without any correspondence with the radio continuum shell of 
G15.4+0.1. A similar behaviour is found between 0 and 41.6~km~s$^{-1}$ and therefore,
the images are not shown. 
In the images corresponding to velocities between 50 and 91.1~km~s$^{-1}$, we observed the
best morphological signature of neutral gas related to the SNR.
Indeed, at $\sim$54~km~s$^{-1}$, the HI distribution appears to be distorted at the position
of the synchrotron emission from G15.4+0.1. Moreover, it is in the 
58.1-91.1~km~s$^{-1}$ range where we see a cavity in the HI emission that closely matches 
G15.4+0.1. In particular, at velocities from 58.1 to 74.6~km~s$^{-1}$, the
HI distribution has an almost complete shell-like morphology 
along the eastern and western peripheries (left and right sides) of the remnant, 
which looks broken toward the 
northeastern border. The presence of this shell is compatible with the central systemic velocity 
$v_{\mathrm{sys}}$= 60~km~s$^{-1}$ derived in C13 for the SNR. The expanding 
shock wave of the SNR on the HI surroundings might be responsible for the creation of the 
observed HI structure. An alternative origin for the HI shell may be a pre-existing stellar wind 
bubble formed for instance by a group of OB stars, one of which could have been 
the SN precursor of G15.4+0.1. In this particular scenario, the brightest synchrotron emission 
observed towards the northwest face of the remnant 
could be where the shock is interacting with the inner wall of the bubble. Furthermore, the fact that 
$^{13}$CO emission was detected on the shell's central face of G15.4+0.1 at approximately 50~km~s$^{-1}$ (C13) 
may indicate that the molecular cloud is outside the front wall of atomic hydrogen; 
perhaps the inner wall is also the place where the synchrotron emission originates. 
Examples of bubbles created by the action of the precursor's stellar winds, 
inside of which evolves a SNR shock wave, are those discovered surrounding the remnants 
Kes~79 \citep{gia09}, 3C~434.1 \citep{fos04}, G106.3+2.7 \citep{kho01}. 
In any case, whether the HI shell is formed by neutral gas accelerated by the SNR shock or by stellar 
winds from the SN progenitor, the X-ray emission detected inside the radio shell of the SNR reinforces the 
idea that the center has been evacuated. 
It is also worth noting that at higher velocities there is no evidence of the naturally expected 
cap-like features associated with an HI shell. The non-detection of possible caps does not imply that these 
structures do not exist. They could be undetected because of confusion
caused by unrelated foreground and background inhomogeneities in the ISM. 
The caps, produced by accelerated HI gas, have been observed either in emission as central concentrations 
projected within the remnant \citep{vel02} or in HI self-absorption in
an expanding stellar wind bubble wall \citep{fos04}. 
At velocities higher than $\sim$91.1~km~s$^{-1}$,
the 21~cm line emission does not correlate with the SNR, and neutral gas is preferentially oriented 
parallel to the Galactic plane. 

In what follows, we derive a number of parameters characterizing the HI shell. For this structure, 
centered at $l$$\sim$$15^{\circ}.42$, $b$$\sim$$0^{\circ}.16$, we assigned a radius 
of $\sim$9$^{\prime}$.8 or $\sim$13.7~pc at 
a distance of 4.8~kpc. If we assume that the HI emission is optically thin,
the integration for velocities between 50 and 91~km~s$^{-1}$,
for which we observed the 
most evident signatures of association between the HI gas and the SNR,  yields an average column 
density $N_{\mathrm{H}}\simeq4\times 10^{20}$~cm$^{-2}$. The total mass of atomic gas 
that forms the shell is about 1.9 $\times$ $10^{3}$~M$_{\odot}$.
This value is in fact an upper limit due to the impossibility of disentangling 
related and unrelated HI gas. If the HI shell was formed by ISM swept up by the SNR shock and assuming that 
before the SN event the total neutral mass was uniformly distributed inside the volume of the HI shell, 
the derived mean density of hydrogen nuclei 
in the ambient environment is $\eta_{0}\sim$7~cm$^{-3}$, which is 
larger than the typical interstellar hydrogen densities 
of $\sim$1~cm$^{-3}$ \citep[averaged over the cold, warm, and hot gas-phase constituents of the ISM,][]{mck77}.
Our result corresponds to a reasonable kinetic energy supplied by the SN explosion into the
the surrounding medium of about 7.6 $\times$ 10$^{48}$~erg.
This value was estimated by adopting an expansion velocity for the HI shell of $\sim$20~km~s$^{-1}$, based on 
the systemic velocity of the neutral gas derived in C13 ($v_{\mathrm{sys}}$$\sim$60~km~s$^{-1}$) and the HI-continuum correspondences
seen between $\sim$50 and 91~km~s$^{-1}$. 
Taking into account errors in the selection of the boundaries for integration, confusion from
background or foreground unrelated HI gas, as well as uncertainties in the determination of the distance, we  derived 
the parameters for the HI shell with 
a mean relative error of 45$\%$.

As a by-product, 
on the basis of the new radio continuum image at 624~MHz and the results derived from the HI data, 
we calculated physical parameters of the remnant G15.4+0.1. 
By combining the ambient density $\eta_{0}\sim$7~cm$^{-3}$, the radius of G15.4+0.1 
$R_{\mathrm{SNR}}$$\sim$7$^{\prime}$.2 ($\sim$10.1~pc at the distance of 4.8~kpc) with the 
velocity of the shock, $v_{\mathrm{sh}}$, 
we can roughly calculate the energy released by the SN
explosion to the ISM from the expression given in the model discussed by \citet{che74},

\begin{equation}
E_{\mathrm{SN}}=5.3\,\times10^{43}\, \eta_{0}^{1.12}\, v_{\mathrm{sh}}^{1.40} \, R_{\mathrm{SNR}}^{3.12}\;\mathrm{erg}
\label{E}
.\end{equation}

To estimate the shock velocity, we used 
$v_{\mathrm{sh}}=66.5\,(\frac{R_{\mathrm{SNR}}}{21.9})^{-2.23}$~km~s$^{-1}$ and obtained 
$v_{\mathrm{sh}}=370$~km~s$^{-1}$ for a shock radius set to $R_{\mathrm{SNR}}\sim$10.1~pc. 
We understand that this value represents a reliable bound for the velocity of the SN shock before
impacting the dense molecular cloud located in front of G15.4+0.1; 
the velocity range of 46-50.3~km~s$^{-1}$ for
the cloud corresponds to a distance of $\sim$4.2~kpc (C13).
To test the robustness of our interpretation, we also estimated the shock velocity 
using  the physical parameters derived for the
$^{13}$CO cloud interacting with SNR~G15.4+0.1, as calculated with the expression
$v_{\mathrm{sh}}=v_{\mathrm{cl}}\,(\eta_{\mathrm{cl}}/\eta_{\mathrm{int}})^{1/2}$, where 
$\eta_{\mathrm{cl}}$ and $v_{\mathrm{cl}}$
are the density and the expansion velocity of cloud, respectively, 
and $\eta_{\mathrm{int}}$ is the intercloud density assumed to be 1~cm$^{-3}$ \citep{dub99}.
In C13, it was demonstrated that the CO cloud interacting with G15.4+0.1 has a density of 
$\sim$1.5 $\times 10^{3}$~cm$^{-3}$ 
and it is expanding at $\sim$10~km~s$^{-1}$. 
In these circumstances, we derived a shock velocity of
$v_{\mathrm{sh}}\simeq380$~km~s$^{-1}$, which agrees within the uncertainties of the method 
with that obtained using HI data. Finally, for an average  shock velocity of 375~km~s$^{-1}$ and
using the ambient density along with the radius of the shell   
derived from our study, we obtain an initial energy of $\sim$2.6 $\times 10^{51}$~erg, 
which agrees with the standard value.

To account for the age of G15.4+0.1, we employed the 
dynamical evolution model of \citet{che74} and obtained 
$t\simeq (\frac{R_{SNR}}{21.9})^{3.23} 10^{5} \simeq$ 8.2 $\times$10$^{3}$~yr, 
where we used a radius for the SNR of $R_{SNR}$$\sim$10.1~pc at a distance of about 4.8~kpc. 
This result differs from that calculated by \citet{abr14} (2.5$\times10^{3}$~yr) based on a Sedov-Taylor 
and the assumption of an ambient density around G15.4+0.1 of 1~cm$^{-3}$. 

\subsection{Modelling the broadband spectrum of G15.4+0.1}
The analysis in \citet{abr14} to determine the origin of the $\gamma$-ray emission in the region 
of G15.4+0.1 excludes a hadronic contribution. In modelling the spectral energy distribution (SED),  
these authors used radio data at 330 and 1400~MHz from the GPS \citep{bro05gps} 
and the Multi-Array Galactic Plane Imaging Survey \citep[MAGPIS,][]{hel06}, respectively, whose fluxes were treated 
as upper limits, X-ray data from \it XMM-Newton\rm, and $\gamma$-ray observations obtained with 
H.E.S.S. In this section, we first demonstrate that, with the available data, 
a hadronic model in which the TeV $\gamma$-ray flux is produced through the interaction of 
accelerated protons with ambient protons, mainly from the molecular cloud located foreground G15.4+0.1,
is still compatible with the broadband spectrum of the G15.4+0.1/HESS~J1818$-$154 system. In addition,
in the light of the new GMRT data, we revisited the leptonic model presented in \citet{abr14} 
to fit the multiwavelength spectral data.

The energy spectrum of the accelerated particles (electrons and protons) in the emission region
is described by a power law with an exponential cut-off,

\begin{equation}
\label{Spec}
\frac{dn_{e,p}}{dE} = K_{e,p}\ E^{-\Gamma_{e,p}} \exp\left(-E/E_{cut_{e,p}} \right),
\end{equation}   

\noindent
where the subscripts \it e \rm and \it p \rm refer to the particle species (electrons or protons), 
and $E$, $K$, $\Gamma$, and $E_{\mathrm{cut}}$ are the particle energy, a normalization 
constant, the spectral index, and the cut-off energy, respectively. 
We refer to \citet{aha10} and references therein for calculation details of the synchrotron emissivity produced 
by electrons. Regarding the inverse Compton (IC) and the non-thermal Bremsstrahlung emission, 
we followed the method given in \citet{jon68} for the former, while the method presented
in  \citet{bar99} was used to model the electron-electron Bremsstrahlung interaction and in
 \citet{koc59} and \citet{stu97} for the  electron-ion Bremsstrahlung process. 
On the other hand, we determine the $\gamma$-ray spectrum due to the neutral pion decays
by using the \it ppfrag \rm program \citep{kac12}. 
In the calculations, the low and high energy parts of the spectrum were separately considered.
For energies of the $\gamma$ rays smaller than 50~GeV, we used the parametrization 
presented in \citet{kam07}, while for larger values of energy we based our calculations in 
the interaction model QGSJET-II-04\footnote{The last version of the QGSJET model was updated
with recent Large Hadron Collider data.} \citep{ost11}.

\begin{figure}[!h]
\centering
\includegraphics[width=8cm]{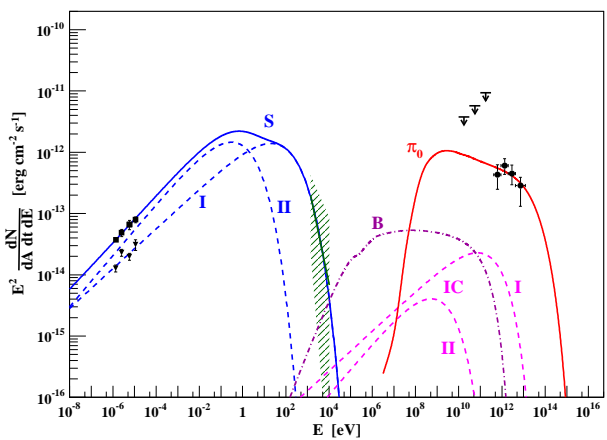}
\caption{Spectral energy distribution of the SNR~G15.4+0.1. 
The filled squares and triangles with error bars correspond to the emission from the whole remnant
and from the interior of the shell (Region I), respectively \citep[data from][and this work]{bro06}.  
The shadowed region encompassing a solid line corresponds to the fitted power law and the one sigma 
region of the X-ray data. The HESS TeV data points are indicated by filled circles\citep{abr14}, while 
the arrows pointing downwards show the upper limits in the \it Fermi\rm-LAT spectrum \citep{ace13}.
The dashed lines labelled I and II mark the spectra from Region I, assuming non-thermal X-ray and radio emission
originating in the interior of G15.4+0.1, and from Region II in the shell of the remnant, respectively.
The modelled spectra from the synchrotron radiation 
by electrons, inverse Compton on the CMB, Bremsstrahlung, and $\gamma$ rays produced by neutral pions 
generated in proton-proton interactions are indicated by the 
curves labelled {\bf S}, {\bf IC}, {\bf B,} and $\bf{\pi_0}$, respectively.
} \label{sed-all}
\end{figure}

\it The XMM-Newton \rm observations show faint X-ray emission, 
about 9~arcmin in size, which 
partially coincides with the size of the HESS~J1818$-$154 source (see Fig.~\ref{multiwave}). 
The origin of the X-rays was considered non-thermal and based on their spatial correspondence
with the $\gamma$ rays, 
an association between them was suggested \citep{abr14}. 
We model the spectrum of G15.4+0.1 from radio to the TeV $\gamma$ rays 
by considering a non-thermal nature of the X-ray emission. This implies that
the whole X-ray flux and part of the radio flux is produced by synchrotron
radiation of electrons in the turbulent magnetic field in the region interior to the SNR shell, 
the PWN proposed in \citet{abr14}. We refer to this zone as Region I.
The remaining part of the radio flux is also generated  by synchrotron radiation, but corresponding
to a different population of electrons that are placed in the shell and  referred to 
hereafter as Region II. Figure~\ref{sed-all} shows the broadband spectrum for the SNR G15.4+0.1/HESS~J1818$-$154 system.
The radio data for Regions I and II include the observations presented in \citet{bro06}
and the new GMRT data at 624 and 1420~MHz.
The spectrum also includes the best fit to the X-ray data using a power law taken from \citet{abr14},
and upper limit $\gamma$-ray fluxes obtained by \it Fermi\rm-LAT \citep{ace13} and H.E.S.S \citep{abr14}. 
Although upper limits on the fluxes at GeV energies are represented, they were not used 
in fitting the spectrum as they  are too far from HESS data. 
The best fit to the radio data corresponding to the 
interior SNR region is shown by a dashed curve labelled I. 
The intensity of the magnetic field and the cut-off energy are fixed during the fitting procedure in order 
to be consistent with the power-law fit to the X-ray data. The normalization constant $K$ 
and the spectral index $\Gamma$ in Eq.~(\ref{Spec}) are taken as free fit parameters. The 
intensity of the magnetic field used is $B^{\,\textrm{I}}=25\ \mu$G and the cut-off energy is 
$\log(E^{\,\textrm{I}}_{cut,e}/\mathrm{eV})=12.83$. These values are chosen in such a way that the inverse 
Compton component has a sub-dominant role at very high energies. It is worth mentioning that scenarios with
larger values of the magnetic field intensity and smaller cut-off energies of the electron 
component, such that 
$(E^{\,\textrm{I}}_{cut,e})^2\times B^{\,\textrm{I}} = 1.1536\times 10^{21}$ eV$^2$ G, also describe 
properly the X-ray data, making the inverse Compton component even smaller. The 
best-fit model for Region I yields a spectral index $\Gamma^{\,\textrm{I}}_{e}=2.35 \pm 0.14$.

We also considered a contribution to the leptonic mechanism from accelerated electrons in the SNR shell. 
The radio flux, corresponding to the shell region, is obtained by fitting the total radio flux,
by using the best fit to the radio data corresponding to the internal region. In this case, we fixed
the cut-off energy and the intensity of the magnetic field to 
$\log(E^{\,\textrm{II}}_{cut,e}/\textrm{eV}) = 11.6$ 
and $B^{\,\textrm{II}}=62.5\ \mu$G, respectively. This choice of the parameters prevents the synchrotron 
flux that originated in the shell region from contributing to the X-ray part of the SED. We note that this 
combination of the two parameters is not unique. Smaller values of the cut-off energy, which are 
consistent with the data, produce smaller values of the high energy flux originated by the inverse Compton process.
The spectral index obtained from the fit is 
$\Gamma^{\,\textrm{II}}_{e}=2.16 \pm 0.08$.            
It is evident that in our model the inverse Compton process alone cannot explain  the 
emission at TeV energies from Region I or II.
Figure~\ref{sed-all} also includes the spectrum corresponding to the Bremsstrahlung radiation. In this
case, we assumed that only  Region II has enough density to produce a non-negligible flux.
The calculation was done by fixing the proton 
density, both molecular and atomic, 
to the value $n=6.4$~cm$^{-3}$ averaged over  Region II. We note that, as in the IC model, 
it is possible to reduce even more the Bremsstrahlung effects diminishing the value of the cut-off energy 
of the accelerated electrons of the shell region. 

It is worth mentioning that the magnetic field intensity of the two regions required to describe
the data is much larger than the field corresponding to the interstellar medium, for which 
its typical strength is 1-2~$\mu$G \citep[][and references therein]{fos13}. This is typical  
in hadronic models of the $\gamma$-ray emission \citep[see e.g. ][]{car14}.

We determined the hadronic contribution of the $\gamma$-ray flux resulting from the 
neutral pion decay of accelerated protons (illustrated by a solid line labelled as $\pi^{0}$ 
in Fig.~\ref{sed-all}). 
Since neither the H.E.S.S. data nor the \it Fermi\rm-LAT upper limits are 
enough to constrain the spectral index of the proton component, we used the spectral
index of the electron component fitted for the synchrotron emission 
in Region II, 
that is $\Gamma^{\,\textrm{II}}_{p}=2.16$. 
Taking into account the normalization constant and the cut-off energy as
free fit parameters, we obtained for the proton component 
a cut-off energy $\log(E_{cut,p}/\textrm{eV}) = 14.6 \pm 0.5$.

\begin{figure}[!h]
\centering
\includegraphics[width=8cm]{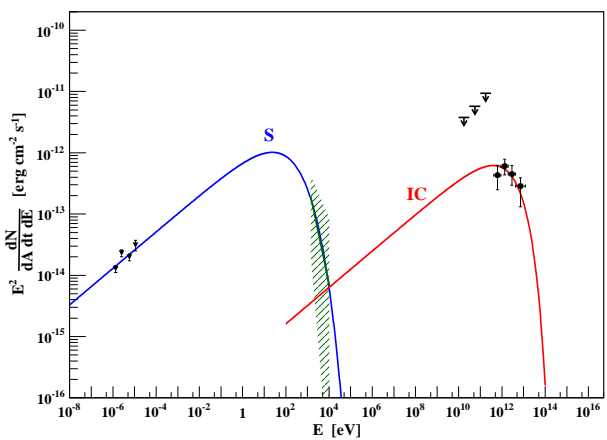}
\caption{Broadband fit to radio (filled triangles) 
and $\gamma$-ray observations obtained with H.E.S.S. (filled circles). 
Upper limits to the flux densities from \it Fermi\rm-LAT (arrows pointing downwards) 
and the best fit to the X-ray data (green hatching) 
are also included. The radio data correspond to the internal region of SNR~G15.4+0.1. The curves labelled 
{\bf S} and {\bf IC} show the model of the synchrotron radiation and inverse Compton 
mechanism on the CMB by accelerated 
electron, respectively.
} 
\label{sed-one-zone}
\end{figure}

On the other hand, we revisited the one-zone model presented in \citet{abr14} replacing the upper 
limits on the radio fluxes considered in that work by reliable flux density measurements
of the radio synchrotron emission in the interior of G15.4+0.1 
(called Region I in the current work). This region is spatially  coincident with the X-ray emission
 attributed by those authors to the same electron population responsible for the $\gamma$-ray emission
through the IC process. In this case as well, a power law with an exponential cut-off was assumed to model 
the spectral energy distribution of electrons.
As was done for the hadronic scenario, we imposed the condition 
$(E_{cut,e})^{2} \times B = 1.1536 \times 10^{21}$~eV$^{2}$~G, 
which ensures an adequate description of the X-ray data. We would like to note that
also in this case  the \it Fermi\rm-LAT upper limits do not constrain the model.
The corresponding spectrum is shown in Fig.~\ref{sed-one-zone}. For this simple model, 
we obtain a spectral index $\Gamma_{e}=2.41 \pm 0.09$, a magnetic field
$B = 3.9 \pm 1.4$~$\mu$G, and a cut-off energy of electrons $\log(E_{cut,e}/\textrm{eV}) = 13.303 \pm 0.071$. 
The fit corresponding to this scenario is as good as the one obtained in \citet{abr14}. However, the 
fitted parameters derived in this work are restricted to smaller intervals. This is due to the radio data 
points of the internal region used to fit the SED.

The observational data is compatible with both scenarios analyzed here; more data is needed, 
especially in the energy region relevant to the \it Fermi\rm-LAT observatory, in order to make progress in 
the understanding of the most important physical processes taking place in this source. In any case, 
it is important to emphasize that a hadronic scenario is compatible with the observational data.   

\section{Conclusions}
We presented new full-synthesis imaging of the SNR~G15.4+0.1 obtained from observations at 624~MHz 
carried out with the GMRT.
We measured an integrated flux density 
$S=8.0 \pm 1.1$~Jy at 624~MHz. 
Based on the combination of our estimate with those previously published, after
bringing all values to the same absolute flux density scale, we derive a spectral index 
$\alpha$=$-0.62 \pm 0.03$ for the whole remnant. The estimated global spectral index is consistent with
measurements for other typical SNR shells \citep{kot06}.

The new GMRT data were used to analyze the correspondences with FIR and X-ray
observations towards the remnant. We found an impressive correlation between the infrared emission
detected by \it Herschel \rm at 70 and 250~$\mu$m and the cloud of $^{13}$CO colliding with the SNR shock
reported in C13. In other regions of G15.4+0.1, however, no IR counterpart is observed to the 
radio emission. On the other hand, from the comparison of the new 624~MHz image with re-processed \it XMM-Newton \rm 
observations between 1 and 8~keV, we can confirm that the synchrotron radio emission surrounds the
low surface brightness X-ray radiation. No counterparts were observed in radio to the five 
point-like sources detected in the \it XMM-Newton \rm field. 
We also conducted a search for a neutron star associated with G15.4+0.1 through time-series
observations performed with the GMRT, which provided negative results.

The analysis of the spatial spectral index variations 
made between existing observations of G15.4+0.1 at 330 and the new GMRT data at 624~MHz 
revealed a radio spectrum steepening from the weak interior to
the brighter periphery of G15.4+0.1.
A similar spectral behaviour is also found in other young SNRs.
The overall distribution of the local spectral radio index 
could be described by diffusive shock acceleration by considering the effect of the ISM gas. 
We did not recognize any distinct feature  in the radio brightness at 624~MHz or in the 
330/624 spectral distribution  that could be considered  a synchrotron  
nebula, powered by the wind of a yet undetected pulsar. 
Furthermore, our search for radio pulsations inside the 
synchrotron shell did not reveal a pulsar, with the upper limits on 
the mean flux density of any associated putative pulsar of 250 and 
300~$\mu$Jy at 624 and 1404~MHz, respectively, towards the centroid of the TeV source. 
In addition, the flux density upper limits that we estimated for any 
putative pulsar morphologically coincident with SNR G15.4+0.1, 
including the five point-like X-ray sources reported by \citet{abr14}, 
were of about 700 and 500~$\mu$Jy at 624 and 1404~MHz, respectively .

Our study of the neutral ambient gas in the direction of G15.4+0.1 unveiled a large 
(about 13.7~pc in radius at a distance of 4.8~kpc) and incomplete
HI shell-type structure. We determined
a  mass for this shell of $1.9 \times 10^{3}$~M$_ {\odot}$ and a kinetic energy 
of $7.6 \times 10^{48}$~erg. On the basis of
this study and the new SNR image at 624~MHz, we estimated that G15.4+0.1 was created as the result of
$2.6 \times 10^{51}$~erg released in a SN explosion that occurred about $8.2 \times 10^{3}$~yr ago. 

Finally, we discussed different models to fit the broadband emission from the 
SNR~G15.4+0.1/HESS~J1818$-$154 system. In particular, we demonstrated that taking into consideration 
the multiwavelength data known to date,
a purely hadronic picture is sufficient 
to account for the $\gamma$-ray data. This is in contrast to the explanation given in \citet{abr14} 
for which the X-ray and TeV emission have the same origin. 
We recognize that more comprehensive 
X-ray and $\gamma$-ray data (especially in the MeV-GeV band) are required 
to improve the constraints on the observed SED of the system and distinguish between leptonic and 
hadronic explanations for the origin of HESS~J1818$-$154.

\begin{acknowledgements}
We wish to acknowledge the referee for his very fruitful comments that improved our work. 
We thank C.~Brogan for supplying us with the VLA 330~MHz image. 
This research has been funded by Argentina
grants ANPCYT-PICT 0795/08 and 0571/11. G.~C. is Member of the Carrera of Investigator Cient\'ifico of
CONICET, Argentina. L.~S. is a Ph.D. Fellow of CONICET, Argentina.  
The GOODS-Herschel data was accessed through the Herschel Database in Marseille 
(HeDaM - http://hedam.lam.fr) operated by CeSAM and hosted by the Laboratoire d'Astrophysique de Marseille. We thank the staff of the GMRT, who have made these observations possible. GMRT is operated by the National Centre for Radio Astrophysics of the Tata Institute of Fundamental Research. This work made 
use of the High Performance Computing facility, funded by grant 12-R\&D-TFR-5.02-0711, at National Centre for Radio Astrophysics of the Tata Institute of Fundamental Research.

\end{acknowledgements}

\bibliographystyle{aa}
 \bibliography{castelletti-astroph}

\begin{thebibliography}{42}
\expandafter\ifx\csname natexlab\endcsname\relax\def\natexlab#1{#1}\fi

\bibitem[{{Abramowski} {et~al.}(2014){Abramowski}, {Aharonian}, {Ait Benkhali},
  {Akhperjanian}, {Ang{\"u}ner}, {Anton}, {Balenderan}, {Balzer}, {Barnacka},
  \& et~al.}]{abr14}
{Abramowski}, A., {Aharonian}, F., {Ait Benkhali}, F., {et~al.} 2014, \aap,
  562, A40

\bibitem[{{Acero} {et~al.}(2013){Acero}, {Ackermann}, {Ajello}, \&
  et~al.}]{ace13}
{Acero}, F., {Ackermann}, M., {Ajello}, M., \& et~al. 2013, \apj, 773, 77

\bibitem[{{Aharonian} {et~al.}(2010){Aharonian}, {Kelner}, \&
  {Prosekin}}]{aha10}
{Aharonian}, F.~A., {Kelner}, S.~R., \& {Prosekin}, A.~Y. 2010, \prd, 82,
  043002

\bibitem[{{Anderson} {et~al.}(2011){Anderson}, {Bania}, {Balser}, \&
  {Rood}}]{and11}
{Anderson}, L.~D., {Bania}, T.~M., {Balser}, D.~S., \& {Rood}, R.~T. 2011,
  VizieR Online Data Catalog, 219, 40032

\bibitem[{{Anderson} \& {Rudnick}(1993)}]{and93}
{Anderson}, M.~C. \& {Rudnick}, L. 1993, \apj, 408, 514

\bibitem[{{Baring} {et~al.}(1999){Baring}, {Ellison}, {Reynolds}, {Grenier}, \&
  {Goret}}]{bar99}
{Baring}, M.~G., {Ellison}, D.~C., {Reynolds}, S.~P., {Grenier}, I.~A., \&
  {Goret}, P. 1999, \apj, 513, 311

\bibitem[{{Brogan} {et~al.}(2005{\natexlab{a}}){Brogan}, {Gaensler}, {Gelfand},
  {Roberts}, {Lazio}, \& {Kassim}}]{bro05gps}
{Brogan}, C.~L., {Gaensler}, B.~M., {Gelfand}, Y., {et~al.} 2005{\natexlab{a}},
  in X-Ray and Radio Connections, ed. L.~O. {Sjouwerman} \& K.~K. {Dyer}, 4

\bibitem[{{Brogan} {et~al.}(2006){Brogan}, {Gelfand}, {Gaensler}, {Kassim}, \&
  {Lazio}}]{bro06}
{Brogan}, C.~L., {Gelfand}, J.~D., {Gaensler}, B.~M., {Kassim}, N.~E., \&
  {Lazio}, T.~J.~W. 2006, \apjl, 639, L25

\bibitem[{{Brogan} {et~al.}(2005{\natexlab{b}}){Brogan}, {Lazio}, {Kassim}, \&
  {Dyer}}]{bro05}
{Brogan}, C.~L., {Lazio}, T.~J., {Kassim}, N.~E., \& {Dyer}, K.~K.
  2005{\natexlab{b}}, \aj, 130, 148

\bibitem[{{Ca{\~n}ellas} {et~al.}(2012){Ca{\~n}ellas}, {Joshi}, {Paredes},
  {Ishwara-Chandra}, {Mold{\'o}n}, {Zabalza}, {Mart{\'{\i}}}, \&
  {Rib{\'o}}}]{cjp12}
{Ca{\~n}ellas}, A., {Joshi}, B.~C., {Paredes}, J.~M., {et~al.} 2012, \aap, 543,
  A122

\bibitem[{{Cardillo} {et~al.}(2014){Cardillo}, {Tavani}, {Giuliani},
  {Yoshiike}, {Sano}, {Fukuda}, {Fukui}, {Castelletti}, \& {Dubner}}]{car14}
{Cardillo}, M., {Tavani}, M., {Giuliani}, A., {et~al.} 2014, \aap, 565, A74

\bibitem[{{Castelletti} {et~al.}(2011){Castelletti}, {Dubner}, {Clarke}, \&
  {Kassim}}]{cas11}
{Castelletti}, G., {Dubner}, G., {Clarke}, T., \& {Kassim}, N.~E. 2011, \aap,
  534, A21

\bibitem[{{Castelletti} {et~al.}(2013){Castelletti}, {Supan}, {Dubner},
  {Joshi}, \& {Surnis}}]{cas13}
{Castelletti}, G., {Supan}, L., {Dubner}, G., {Joshi}, B.~C., \& {Surnis},
  M.~P. 2013, \aap, 557, L15 (C13)

\bibitem[{{Chevalier}(1974)}]{che74}
{Chevalier}, R.~A. 1974, \apj, 188, 501

\bibitem[{{Cordes} \& {Lazio}(2002)}]{cor02}
{Cordes}, J.~M. \& {Lazio}, T.~J.~W. 2002, ArXiv Astrophysics e-prints,
  astro-ph/0207156

\bibitem[{{Dubner} {et~al.}(1999){Dubner}, {Giacani}, {Reynoso}, {Goss},
  {Roth}, \& {Green}}]{dub99}
{Dubner}, G., {Giacani}, E., {Reynoso}, E., {et~al.} 1999, \aj, 118, 930

\bibitem[{{Foster} {et~al.}(2013){Foster}, {Kothes}, \& {Brown}}]{fos13}
{Foster}, T., {Kothes}, R., \& {Brown}, J.~C. 2013, \apjl, 773, L11

\bibitem[{{Foster} {et~al.}(2004){Foster}, {Routledge}, \& {Kothes}}]{fos04}
{Foster}, T., {Routledge}, D., \& {Kothes}, R. 2004, \aap, 417, 79

\bibitem[{{Giacani} {et~al.}(2009){Giacani}, {Smith}, {Dubner}, {Loiseau},
  {Castelletti}, \& {Paron}}]{gia09}
{Giacani}, E., {Smith}, M.~J.~S., {Dubner}, G., {et~al.} 2009, \aap, 507, 841

\bibitem[{{Griffin} {et~al.}(2010){Griffin}, {Abergel}, {Abreu}, {Ade},
  {Andr{\'e}}, {Augueres}, {Babbedge}, {Bae}, {Baillie}, {Baluteau}, {Barlow},
  {Bendo}, {Benielli}, {Bock}, {Bonhomme}, {Brisbin}, {Brockley-Blatt},
  {Caldwell}, {Cara}, {Castro-Rodriguez}, {Cerulli}, {Chanial}, {Chen},
  {Clark}, {Clements}, {Clerc}, {Coker}, {Communal}, {Conversi}, {Cox},
  {Crumb}, {Cunningham}, {Daly}, {Davis}, {de Antoni}, {Delderfield}, {Devin},
  {di Giorgio}, {Didschuns}, {Dohlen}, {Donati}, {Dowell}, {Dowell}, {Duband},
  {Dumaye}, {Emery}, {Ferlet}, {Ferrand}, {Fontignie}, {Fox}, {Franceschini},
  {Frerking}, {Fulton}, {Garcia}, {Gastaud}, {Gear}, {Glenn}, {Goizel},
  {Griffin}, {Grundy}, {Guest}, {Guillemet}, {Hargrave}, {Harwit}, {Hastings},
  {Hatziminaoglou}, {Herman}, {Hinde}, {Hristov}, {Huang}, {Imhof}, {Isaak},
  {Israelsson}, {Ivison}, {Jennings}, {Kiernan}, {King}, {Lange}, {Latter},
  {Laurent}, {Laurent}, {Leeks}, {Lellouch}, {Levenson}, {Li}, {Li},
  {Lilienthal}, {Lim}, {Liu}, {Lu}, {Madden}, {Mainetti}, {Marliani}, {McKay},
  {Mercier}, {Molinari}, {Morris}, {Moseley}, {Mulder}, {Mur}, {Naylor},
  {Nguyen}, {O'Halloran}, {Oliver}, {Olofsson}, {Olofsson}, {Orfei}, {Page},
  {Pain}, {Panuzzo}, {Papageorgiou}, {Parks}, {Parr-Burman}, {Pearce},
  {Pearson}, {P{\'e}rez-Fournon}, {Pinsard}, {Pisano}, {Podosek}, {Pohlen},
  {Polehampton}, {Pouliquen}, {Rigopoulou}, {Rizzo}, {Roseboom}, {Roussel},
  {Rowan-Robinson}, {Rownd}, {Saraceno}, {Sauvage}, {Savage}, {Savini},
  {Sawyer}, {Scharmberg}, {Schmitt}, {Schneider}, {Schulz}, {Schwartz},
  {Shafer}, {Shupe}, {Sibthorpe}, {Sidher}, {Smith}, {Smith}, {Smith},
  {Spencer}, {Stobie}, {Sudiwala}, {Sukhatme}, {Surace}, {Stevens}, {Swinyard},
  {Trichas}, {Tourette}, {Triou}, {Tseng}, {Tucker}, {Turner}, {Vaccari},
  {Valtchanov}, {Vigroux}, {Virique}, {Voellmer}, {Walker}, {Ward}, {Waskett},
  {Weilert}, {Wesson}, {White}, {Whitehouse}, {Wilson}, {Winter}, {Woodcraft},
  {Wright}, {Xu}, {Zavagno}, {Zemcov}, {Zhang}, \& {Zonca}}]{gri10}
{Griffin}, M.~J., {Abergel}, A., {Abreu}, A., {et~al.} 2010, \aap, 518, L3

\bibitem[{{Helfand} {et~al.}(2006){Helfand}, {Becker}, {White}, {Fallon}, \&
  {Tuttle}}]{hel06}
{Helfand}, D.~J., {Becker}, R.~H., {White}, R.~L., {Fallon}, A., \& {Tuttle},
  S. 2006, \aj, 131, 2525

\bibitem[{{Hines} {et~al.}(2004){Hines}, {Rieke}, {Gordon}, {Rho}, {Misselt},
  {Woodward}, {Werner}, {Krause}, {Latter}, {Engelbracht}, {Egami}, {Kelly},
  {Muzerolle}, {Stansberry}, {Su}, {Morrison}, {Young}, {Noriega-Crespo},
  {Padgett}, {Gehrz}, {Polomski}, {Beeman}, \& {Haller}}]{hin04}
{Hines}, D.~C., {Rieke}, G.~H., {Gordon}, K.~D., {et~al.} 2004, \apjs, 154, 290

\bibitem[{{Hofverberg} {et~al.}(2011){Hofverberg}, {Chaves}, {M\'{e}hault}, \&
  {de Naurois}}]{hof11}
{Hofverberg}, P., {Chaves}, R.~C.~G., {M\'{e}hault}, J., \& {de Naurois}, M.
  2011, in International Cosmic Ray Conference, Vol.~7, International Cosmic
  Ray Conference, 247

\bibitem[{{Jones}(1968)}]{jon68}
{Jones}, F.~C. 1968, Physical Review, 167, 1159

\bibitem[{{Kachelrie{\ss}} \& {Ostapchenko}(2012)}]{kac12}
{Kachelrie{\ss}}, M. \& {Ostapchenko}, S. 2012, \prd, 86, 043004

\bibitem[{{Kamae} {et~al.}(2007){Kamae}, {Karlsson}, {Mizuno}, {Abe}, \&
  {Koi}}]{kam07}
{Kamae}, T., {Karlsson}, N., {Mizuno}, T., {Abe}, T., \& {Koi}, T. 2007, \apj,
  662, 779

\bibitem[{{Katz-Stone} \& {Rudnick}(1997)}]{kat97}
{Katz-Stone}, D.~M. \& {Rudnick}, L. 1997, \apj, 479, 258

\bibitem[{{Koch} \& {Motz}(1959)}]{koc59}
{Koch}, H.~W. \& {Motz}, J.~W. 1959, Reviews of Modern Physics, 31, 920

\bibitem[{{Kothes} {et~al.}(2006){Kothes}, {Fedotov}, {Foster}, \&
  {Uyan{\i}ker}}]{kot06}
{Kothes}, R., {Fedotov}, K., {Foster}, T.~J., \& {Uyan{\i}ker}, B. 2006, \aap,
  457, 1081

\bibitem[{{Kothes} {et~al.}(2001){Kothes}, {Uyaniker}, \& {Pineault}}]{kho01}
{Kothes}, R., {Uyaniker}, B., \& {Pineault}, S. 2001, \apj, 560, 236

\bibitem[{{McClure-Griffiths} {et~al.}(2005){McClure-Griffiths}, {Dickey},
  {Gaensler}, {Green}, {Haverkorn}, \& {Strasser}}]{mcc05}
{McClure-Griffiths}, N.~M., {Dickey}, J.~M., {Gaensler}, B.~M., {et~al.} 2005,
  \apjs, 158, 178

\bibitem[{{McKee} \& {Ostriker}(1977)}]{mck77}
{McKee}, C.~F. \& {Ostriker}, J.~P. 1977, \apj, 218, 148

\bibitem[{{Ostapchenko}(2011)}]{ost11}
{Ostapchenko}, S. 2011, \prd, 83, 014018

\bibitem[{{Perley} \& {Butler}(2013)}]{per13}
{Perley}, R.~A. \& {Butler}, B.~J. 2013, \apjs, 204, 19

\bibitem[{{Sandberg} \& {Sollerman}(2009)}]{san09}
{Sandberg}, A. \& {Sollerman}, J. 2009, \aap, 504, 525

\bibitem[{{Slane} {et~al.}(2008){Slane}, {Helfand}, {Reynolds}, {Gaensler},
  {Lemiere}, \& {Wang}}]{sla08}
{Slane}, P., {Helfand}, D.~J., {Reynolds}, S.~P., {et~al.} 2008, \apjl, 676,
  L33

\bibitem[{{Sturner} {et~al.}(1997){Sturner}, {Skibo}, {Dermer}, \&
  {Mattox}}]{stu97}
{Sturner}, S.~J., {Skibo}, J.~G., {Dermer}, C.~D., \& {Mattox}, J.~R. 1997,
  \apj, 490, 619

\bibitem[{{Supan} {et~al.}(2014){Supan}, {Castelletti}, {Dubner}, {Surnis}, \&
  {Joshi}}]{sup14}
{Supan}, L., {Castelletti}, G., {Dubner}, G., {Surnis}, M.~P., \& {Joshi},
  B.~C. 2014, Boletin de la Asociacion Argentina de Astronomia La Plata
  Argentina, 56, 315

\bibitem[{{Turner} {et~al.}(2001){Turner}, {Abbey}, {Arnaud}, {Balasini},
  {Barbera}, {Belsole}, {Bennie}, {Bernard}, {Bignami}, {Boer}, {Briel},
  {Butler}, {Cara}, {Chabaud}, {Cole}, {Collura}, {Conte}, {Cros}, {Denby},
  {Dhez}, {Di Coco}, {Dowson}, {Ferrando}, {Ghizzardi}, {Gianotti}, {Goodall},
  {Gretton}, {Griffiths}, {Hainaut}, {Hochedez}, {Holland}, {Jourdain},
  {Kendziorra}, {Lagostina}, {Laine}, {La Palombara}, {Lortholary}, {Lumb},
  {Marty}, {Molendi}, {Pigot}, {Poindron}, {Pounds}, {Reeves}, {Reppin},
  {Rothenflug}, {Salvetat}, {Sauvageot}, {Schmitt}, {Sembay}, {Short},
  {Spragg}, {Stephen}, {Str{\"u}der}, {Tiengo}, {Trifoglio}, {Tr{\"u}mper},
  {Vercellone}, {Vigroux}, {Villa}, {Ward}, {Whitehead}, \& {Zonca}}]{tur01}
{Turner}, M.~J.~L., {Abbey}, A., {Arnaud}, M., {et~al.} 2001, \aap, 365, L27

\bibitem[{{Vel{\'a}zquez} {et~al.}(2002){Vel{\'a}zquez}, {Dubner}, {Goss}, \&
  {Green}}]{vel02}
{Vel{\'a}zquez}, P.~F., {Dubner}, G.~M., {Goss}, W.~M., \& {Green}, A.~J. 2002,
  \aj, 124, 2145

\bibitem[{{Zajczyk} {et~al.}(2012){Zajczyk}, {Gallant}, {Slane}, {Reynolds},
  {Bandiera}, {Gouiff{\`e}s}, {Le Floc'h}, {Comer{\'o}n}, \& {Koch
  Miramond}}]{zaj12}
{Zajczyk}, A., {Gallant}, Y.~A., {Slane}, P., {et~al.} 2012, \aap, 542, A12

\bibitem[{{Zharikov} {et~al.}(2013){Zharikov}, {Zyuzin}, {Shibanov}, \&
  {Mennickent}}]{zha13}
{Zharikov}, S.~V., {Zyuzin}, D.~A., {Shibanov}, Y.~A., \& {Mennickent}, R.~E.
  2013, \aap, 554, A120

\end{thebibliography}

\end{document}